\documentstyle[preprint,eqsecnum,aps,prd,floats,epsf]{revtex}

\makeatletter
\ifx\every@math@size\undefined\else
  \let\old@expast\@expast
  \def\@expast#1{\old@expast{#1}\let\@tempa\reserved@a}
\fi
\makeatother

\def\Re{\;{\cal R}\!e}
\def\Im{\;{\cal I}\!m}

\begin{document}

\preprint{\font\fortssbx=cmssbx10 scaled \magstep2
\hbox to \hsize{
\hbox{\fortssbx University of Wisconsin - Madison}
\hfill$\vcenter{\hbox{\bf MADPH-99-1126} 
                \hbox{December 1999}}$}}

\title{\vspace*{.5in}
The Nucleon Sigma Term from  Threshold Parameters}

\author{M.G. Olsson}

\address{Physics Department, University of Wisconsin, Madison, WI 53706}

\maketitle

\begin{abstract}
A new sum rule determines the nucleon sigma term by the $\pi N$ elastic scattering threshold parameters alone.  This ``threshold" value of $\sigma(2\mu^2)$ provides an independent check of existing values obtained from global dispersive analyses, and also provides a realistic error assignment. Employing a recent analysis of low energy ($T_\pi < 100$~MeV) $\pi N$ data our sum rule implies $\sigma(2\mu^2) = 71\pm 9$~MeV.
\end{abstract}

\thispagestyle{empty}
\newpage

\section{Introduction}

The sigma term is a fundamental scale of chiral physics\cite{weinberg}, a measure of chiral symmetry breaking. For decades there has been a lively discussion between ``large" sigma term advocates and ``small" ones\cite{osyp,reya,gasser,scadron,hohler,dominguez,altar,olsson}. 
We propose here a new evaluation method which is independent of global phase shift analyses. Our expression depends only on threshold parameters. Recent experiments allow phase shifts to be determined using only data near threshold\cite{woolcock}. When combined with the present sum rule, a truly local sigma term value is obtained together with a realistic error.

There are two basic methods for determining the sigma term. In the framework of QCD the expected chiral symmetry-breaking term is proportional to the bare quark mass and transforms as the $(3\bar3)+(\bar33)$ representation of $\rm SU(3)\times SU(3)$. This is equivalent to the original (pre-QCD) GMOR model\cite{gell-mann}. Using a conventional ratio for quark masses, ${m_u+m_d\over 2m_s} \sim {1\over25}$, and assuming the nucleon has no strangeness content, the ``theoretical" sigma term comes out to be about 25~MeV\cite{reya,dominguez}. With higher order corrections and strangeness the predicted sigma term might rise to about 40~MeV\cite{gasser3}. 

The second method employs the ``on-shell theorem"\cite{osyp}, where the sigma term is related to the $\pi N$ scattering amplitude at the ``Cheng-Dasher" point. This method historically yields a larger value of about 65~MeV\cite{hohler}. This method is used here and is described in some detail below. It is the apparent discrepancy between the ``theoretical" first method and the ``experimental" second method that makes the $\pi N$ sigma term so interesting.

The soft pion chiral constraints on the crossing even $\pi N{\to}\pi N$ amplitude $A^{(+)}(\nu, t, q^2, q'^2)$ are most succinctly expressed in terms of
\begin{equation}
\overline D(\nu_B, q^2, q'^2) \equiv A^{(+)} (\nu=0,\ t,\ q^2, q'^2) - {g(q^2) \, g(q'^2) \over M} \,,
\end{equation}
where $t = -(q-q')^2$, $4M\nu_B = t+q^2+q'^2$, $4M\nu = s-u$, $g$ is the renormalized $\pi NN$ coupling constant ${g^2(-\mu^2)\over 4\pi} \equiv {g^2\over 4\pi} \simeq 14$, and $M$ and $\mu$ are the nucleon and charged pion masses.
The Adler PCAC consistency condition\cite{adler} requires that if one pion mass is soft,
\begin{equation}
\overline D (\nu_B = 0,\ q^2 = -\mu^2,\ q'^2=0) =   \overline D (\nu_B = 0,\ q^2 = 0,\ q'^2 = -\mu^2 ) = 0 \,.\label{eq:adler}
\end{equation}
When both pions are soft, the chiral limit predicts that $\overline D(0,0,0) = 0$
and the sigma term is defined as the chiral symmetry breaking in this limit\cite{weinberg,osyp,dominguez,altar}
\begin{equation}
{\sigma(0)\over F_\pi^2} \equiv - \overline D (\nu_B=0,\ q^2=0,\ q'^2=0) \,, \label{eq:sigma}
\end{equation}
where $F_\pi \simeq 93$~MeV. 

The sigma term can be evaluated accurately in terms of the physical (on-shell) amplitude using the well-known ``on-shell theorem"\cite{osyp,dominguez}. Since this result plays a central role in this, and other, determinations of the sigma term, a brief review of its origin is warranted. The amplitude $\overline D(\nu_B=0, q^2, q'^2)$ is assumed to be slowly varying in $q^2$ and $q'^2$ and hence
\begin{equation}
\overline D(0, q^2, q'^2) = -{\sigma(t)\over F_\pi^2} \left[1+{q^2+q'^2\over \mu^2}\right] + R(0, q^2, q'^2)\,, \label{eq:amp}
\end{equation}
where $R(\nu_B, q^2, q'^2)$ contains all non-linear dependence and this remainder function satisfies
\begin{equation}
R(0,0,0) = R(0, -\mu^2, 0) = R(0, 0, -\mu^2) = 0 \,. \label{eq:R=0}
\end{equation}
The amplitude $\overline D(0, q^2, q'^2)$ in (\ref{eq:amp}) clearly satisfies\cite{eq1.4} the the Adler condition (\ref{eq:adler}) and the sigma term definition (\ref{eq:sigma}). The remainder amplitude $R$ is small and will be evaluated by considering the nearby singular contributions primarily due to the $\Delta(1232)$. The on-shell theorem follows by taking $q^2 = q'^2 = -\mu^2$ (and hence $t=2\mu^2$),
\begin{equation}
F_\pi^2 \,\overline D(t=2\mu^2) = \sigma(2\mu^2) - \Delta_R \label{eq:on-shell}
\end{equation}
where
\begin{equation}
\Delta_R = -F_\pi^2   R(0, -\mu^2, -\mu^2) 
\end{equation}
and by definition $R$ is of order $\mu^4$.

A simple model of the $\Delta(1230)$ contribution to $A^{(+)}(\nu=0,\ \nu_B=0,\ q^2, q'^2)$ is obtained through the successful effective Lagrangian $\Delta N\pi$ coupling\cite{olsson2} 
\begin{equation}
L = g_\Delta \overline \Delta_{\mu\,} \Theta_{\mu\nu}\, N \partial_\nu\, \phi + \rm h.c.\;, \quad \Theta_{\mu\nu} = \delta_{\mu\nu} -\left({1\over2}+Z\right) 
\gamma_\mu\gamma_\nu\;,
\end{equation}
yielding
\begin{equation}
R(\nu_B=0,\ q^2, q'^2)  = {2g_\Delta^2 (2M_\Delta + M) q^2 q'^2 \over 9M_\Delta^2 (M_\Delta^2 - M^2)} \,,
\end{equation}
which automatically satisfies the conditions (\ref{eq:R=0}).
Using $g_\Delta^2/4\pi \simeq 0.28\mu^{-2}$\cite{olsson2} we obtain
\begin{equation}
R(\nu_B=0, -\mu^2, -\mu^2) = 7.5\times10^{-3}\mu^{-1} \,, \label{eq:foo}
\end{equation}
which by (\ref{eq:on-shell}) is equivalent to
\begin{equation}
\sigma(2\mu^2) = \left[F_\pi^2\, \overline D(2\mu^2)\right] - (0.46\rm\ MeV)\,. \label{eq:sigmaNN}
\end{equation}

 The only uncertainty arises from the model dependence of $R$.  Several evaluations\cite{osyp,gasser,altar,bernard} agree on the result given in (\ref{eq:foo}) and under very weak assumptions\cite{bernard} that $|\Delta_R| < 2$~MeV and is negligible compared to the sigma term.

In this paper a new threshold method for finding $\sigma(2\mu^2)$ is developed. The important question of determining $\sigma(0)$ is not addressed. In the literature there is some variation in results for the quantity
\begin{equation}
\Delta_\sigma = \sigma(2\mu^2) - \sigma(0) \,.
\end{equation}
A recent one-loop ChPT estimate\cite{borasoy} of the nucleon scalar form factor implies $\Delta_\sigma \sim 5$--7~MeV whereas a dispersive estimate\cite{gasser2} gives $\Delta_\sigma \sim 15$~MeV.

\section{The dispersive sum rule}

As outlined in the introductory section, the key quantity to evaluate is the on-shell amplitude $\overline D(t=2\mu^2) = A^{(+)} (\nu=0,\ t = 2\mu^2) - g^2/M$. Our evaluation involves the difference between two subtracted fixed-$t$ dispersion relations. Both start from the usual fixed momentum transfer dispersion relation\cite{ham,hohler2} for $D^{(+)} (\nu, t) = A^{(+)}(\nu, t) + \nu B^{(+)}(\nu, t)$,
\begin{equation}
\Re D^{(+)}(\nu, t) = -{g^2\over M} {\nu^2\over \nu^2-\nu_B^2} + {2\over\pi}P \int_{\nu_T}^{\infty} {\nu' d\nu'\over\nu'^2-\nu^2} \Im D^{(+)} (\nu', t) \,,
\label{eq:dispersion}
\end{equation}
where $\nu_T = \mu + t/4M$.

\subsection{Initial sigma term sum rule}

From (\ref{eq:dispersion})  we find by evaluating at $\nu=0$ and $\nu_T$ and subtracting,
\begin{equation}
\overline D(t) = D^{(+)}(\nu_T, t) - {g^2\over M} {\nu_B^2\over\nu_T^2-\nu_B^2} - {2\over\pi} \int_{\nu_T}^{\infty} {d\nu\over\nu(\nu^2-\nu_T^2)} \Im D^{(+)}(\nu, t) \,.
\end{equation}
The PV singularity is not present since $\Im D$ vanishes at $\nu=\nu_T$. We now fix $t=2\mu^2$ and change variable to $d\nu = d\omega = kdk/\omega$, where $k$ is the lab pion momentum $k^2 = \omega^2-\mu^2$, to obtain
\begin{equation}
\overline D(2\mu^2) =  D^{(+)}(\nu_T, 2\mu^2) - {2\nu_T^2\over\pi}  \int_{0}^\infty 
{dk (\omega+\mu) \, \Im D^{(+)} (k, 2\mu^2)/k \over \omega (\omega+\mu r) (\omega + \mu + 2\mu r)} \,,   \label{eq:F2mu}
\end{equation}
where now $\nu_T = \mu(1+r)$ and
\begin{equation}
r = \mu/2M \simeq 0.0744 \,. \label{eq:r}
\end{equation}
The Born term vanishes since $\nu_B = 0$ at $t= 2\mu^2$. Although the integral converges at both limits we explicitly remove the threshold parameters by adding and subtracting the quantity
\begin{equation}
\lim_{k\to0} {\Im D^{(+)} (\omega, 2\mu^2)\over k} \equiv \lambda_2 \,,
\label{eq:lim}
\end{equation}
yielding\cite{coeff}
\begin{eqnarray}
&& \mu \overline D(2\mu^2) = \mu D^{(+)}(\nu_T, 2\mu^2) - 1.047 \mu^2\lambda_2 - I_0 \,, \label{eq:sumrule}\\[2ex]
&& I_0 = {2\mu^3 (1+r)^2\over \pi} \int_0^\infty {dk (\omega+\mu)\over \omega (\omega +\mu r) (\omega + \mu + 2\mu r)} \left[ {\Im D^{(+)} (\omega, 2\mu^2) \over k} - \lambda_2 \right] \,, \label{eq:I_0}
\end{eqnarray}
where  we note that each term in (\ref{eq:sumrule}) is dimensionless.

\subsection{Forward dispersive identity}

Returning to (\ref{eq:dispersion}) but now with momentum transfer $t$ fixed at $t=0$ we evaluate at threshold $(\nu=\omega=\mu)$ and above threshold ($\omega>\mu$) and take the difference,
\begin{equation}
{1\over k^2} \left[ \Re D^{(+)}(\omega, 0) - D^{(+)}(\mu, 0) \right] = {g^2\over M} {\omega_B^2\over (\omega^2-\omega_B^2) (\mu^2-\omega_B^2)} + {2\over\pi}P \int_0^\infty {dk'\over k'^2-k^2} {\Im D^{(+)}(\omega', 0)\over k'} \,, \label{eq:diff}
\end{equation}
where $\omega_B = -\mu r$.
To remove the PV singularity we use the identity
\begin{equation}
P\int_0^\infty {dk'\over k'^2-k^2} = 0 \ ; \ \ k^2 > 0
\end{equation}
and the definitions
\begin{eqnarray}
&&\lim_{k\to 0+} {\Im D^{(+)}(\omega, 0) \over k} \equiv \lambda_0 \,,\\[2ex]
&&\lim_{k\to 0+} {\Re D^{(+)}(\omega, 0) - D^{(+)}(\mu, 0)\over k^2} \equiv K\,.
\end{eqnarray}
With the above definitions, and for $k\ll\mu$, Eq.~(\ref{eq:diff}) becomes
\begin{equation}
 K - {g^2\over \mu^2 M} \left( r\over 1-r^2\right)^2 - {2\over\pi} \int_0^\infty {dk\over k^2}\left[ {\Im D^{(+)}(\omega,0)\over k} - \lambda_0 \right] = 0 \,. \label{eq:ident}
\end{equation}
The integrand vanishes at threshold.

\subsection{The final sum rule}

It might be noted that in the relativistic regime $(\omega\simeq k)$ the $\mu \overline D(2\mu^2)$ integrand in (\ref{eq:I_0}) and $\mu\nu_T^2 K$ in (\ref{eq:ident}) are similar. To obtain our final sum rule we subtract $\mu\nu_T^2 = \mu^3(1+r)^2$ times the identity (\ref{eq:ident}) from the $\mu \overline D(2\mu^2)$ sum rule (\ref{eq:sumrule}), giving
\begin{eqnarray}
\mu \overline D(2\mu^2) &=& \mu \left[ D^{(+)}(\nu_T, 2\mu^2) - \nu_T^2 K \right] - 1.047 \mu^2 \lambda_2 + {\mu g^2\over M} \left(r\over 1-r^2\right)^2 - I\,, \label{eq:final}
\\
\noalign{\hbox{where}}
I &=& {2\mu^3\over\pi} (1+r)^2 \int_0^\infty dk \Biggl\{ (\omega + \mu)  {\left[\Im \, D^{(+)}(\omega, 2\mu^2)/k - \lambda_2\right] \over \omega(\omega+ \mu r) (\omega + \mu + 2\mu r)} \nonumber\\
&&\hskip1.7in{} - {1\over k^2} \left[ \Im\, D^{(+)}(\omega, 0)/k - \lambda_0 \right] \Biggr\} \,. \label{eq:I}
\end{eqnarray}
The various quantities $D^{(+)}(\nu_T, 2\mu^2)$, $K$, $\lambda_2$ and $\lambda_0$ can be evaluated in terms of threshold parameters.

\section{Threshold Expansions}

The relation of the invariant amplitude to the partial-wave scattering amplitudes and phase shifts is standard. Using \cite{ham,hohler2} we have
\begin{equation}
{1\over 4\pi} D^{(+)}(\nu, t) = {W\over M} (f_1+f_2) + {t\over 4MQ^2} \biggl[ E(f_1+f_2) - M(f_1-f_2) \biggr] \,,  \label{eq:p-wave}
\end{equation}
where $2M\omega = W^2-M^2-\mu^2$, the c.m.\ momentum is $Q=Mk/W$ and the nucleon c.m.\ energy $E=\sqrt{M^2+Q^2}$. The quantities $f_1$ and $f_2$ are defined by
\begin{eqnarray}
f_{1} &=& \sum_{\ell=0}^\infty f_{\ell+}^{(+)} P'_{\ell +1}(x) - f_{\ell-}^{(+)} P'_{\ell-1}(x) \,.\\
f_2 &=& \sum_{\ell=1}^\infty (f_{\ell-}^{(+)} - f_{\ell+}^{(+)}) P'_\ell(x) \,.
\end{eqnarray}
Here $P'_\ell$ are derivatives of Legendre polynomials with $t = -2Q^2(1-x)$. The above partial wave amplitudes $f_{\ell\pm}^{(+)}$ are the crossing-even combination of the isospin amplitudes $f_{\ell\pm}^{(I)}$,
\begin{equation}
f_{\ell\pm}^{(+)} = {1\over 3} f_{\ell\pm}^{(1/2)} + {2\over3} f_{\ell\pm}^{(3/2)}  \label{eq:iso}
\end{equation}
and the isospin partial-wave amplitudes are
\begin{equation}
f_{\ell\pm}^{(I)} = {1\over 2iQ} \left(\eta e^{2i\delta} - 1\right)_{ I,\,\ell\pm} \ \mathop{\longrightarrow}\limits_{\rm elastic} \ \left(e^{i\delta} \sin\delta\over Q\right)_{ I,\,\ell\pm}  \,. \label{eq:ipwa}
\end{equation}
Close to threshold we have
\begin{eqnarray}
\delta_{I,\,\ell\pm} &\longrightarrow& (a_{\ell\pm}^{(I)}) Q^{2\ell+1} \;, \\[2ex]
{W\over M+\mu} \Re f_{0+}^{(+)} &\longrightarrow& a_{0+}^{(+)} + {1\over3} C^{(+)} k^2 \,. \label{eq:expansion}
\end{eqnarray}
The latter ($s$-wave expansion) is motivated by dispersion relations and the the static model where $C^{(+)}$ is energy independent\cite{ham}.
Using the above threshold expansion, the quantities needed in the sum rule are
\begin{equation}
{1\over4\pi} D^{(+)} (\nu_T, 2\mu^2) = \sum_{\ell=0}^\infty \left(\beta_{\ell-} a_{\ell-}^{(+)} + \beta_{\ell+} a_{\ell+}^{(+)} \right) \mu^{2\ell}\,,
\label{eq:goo}
\end{equation}
where the first few factors $\beta_{\ell\pm}$ are given in Table~\ref{tab:betas} and the crossing-even scattering lengths are expressed, as in (\ref{eq:iso}), in terms of the isospin scattering lengths.
%
%
Corresponding expansions for the remaining quantities appearing in the sum rule (\ref{eq:sumrule}) are
\begin{equation}
{\mu^2\over4\pi}K = (1+2r)^{-1} (a_{1-}^{(+)} + 2a_{1+}^{(+)} ) + {1\over3} (1+2r)\mu C^{(+)} \,.\label{eq:boo}
\end{equation}

\begin{table}[h]
\begin{center}
\begin{minipage}{4.3in}
\caption[]{Proportionality constants between scattering lengths and the threshold invariant amplitude evaluated at $t=2\mu^2$ as defined in (\ref{eq:goo}).
\label{tab:betas}}
\def\arraystretch{1.2}
\begin{tabular}{@{\quad}ccc@{\quad}}
$\ell$& $\beta_{\ell-}$& $\beta_{\ell+}$\\
\hline
0& 0 & $(1+r)^2$\\
1& 1& $3(1+r)^2-1$\\
2& 3& ${15\over2}(1+r)^2-3$\\
3& ${15\over2}$& ${35\over2}(1+r)^2-{15\over2}$\\
4& ${35\over2}$& ${315\over8} (1+r)^2 - {35\over2}$\\
5& ${315\over 8}$& ${693\over8}(1+r)^2 - {315\over8}$\\
6& ${698\over3}$& ${3465\over 16}(1+r)^2 - {693\over8}$
\end{tabular}
\end{minipage}
\end{center}
\end{table}

Finally, we evaluate the two imaginary threshold quantities $\lambda_0$ and $\lambda_2$,
\begin{equation}
\lambda_{0,2} = \lim_{k\to 0} {\Im D^{(+)}(\nu_r, \, t=0{\rm\ or\ }2\mu^2) \over k} \,.
\end{equation}
In each case only the $\ell=0$ partial wave enters and hence only $f_1$ contributes to $D^{(+)}$. Using (\ref{eq:ipwa}) and $QW=kM$ we find
\begin{equation}
{1\over 4\pi}\lambda_0 = {1\over3} \left( a_{0+}^{(1/2)} \right)^2 + {2\over3} \left( a_{0+}^{(3/2)} \right)^2 \,.  \label{eq:lambda0}
\end{equation}
and
\begin{equation}
\lambda_2 = {(1+r)^2\over 1+2r}\lambda_0 = 1.036\lambda_0 \,. \label{eq:lambda2}
\end{equation}
We note that $\lambda_2$ is the same as $\lambda_0$ to better than 4\%.
It should be noted that the amplitude combination $D = A+\nu B$ is unique in giving a threshold expansion without $p$-wave effective ranges in $K$. As one sees from (\ref{eq:p-wave}) the $1/Q^2$ term vanishes at $t=0$.

\section{Evaluation}

In order to evaluate the final sum rule (\ref{eq:final}) we first consider the integral (\ref{eq:I}). The integration is straightforward in terms of the partial wave amplitude (\ref{eq:p-wave}) once the phase shifts and absorption factors are given. We use the very convenient VPI/GW partial wave analysis On Line\cite{arndt}. In Fig.~1 we show the integrand of (\ref{eq:I}) with thirteen crossing-even partial waves $\ell\pm=0+, 1\pm, 2\pm, 3\pm, 4\pm, 5\pm, 6\pm$. Large cancellation between the two parts of the integrand reduce it by nearly an order of magnitude. It is evident that the $\Delta(1230)$ resonance dominates and the integrated result for $I$ is
\begin{equation}
I \simeq 0.21\pm0.02 \,.   \label{eq:continuum}
\end{equation}
The error is less than 10\% and is negligible compared to other errors.

Since the partial wave analysis extends from threshold to about $k=2$~GeV some estimate must be made for the contribution above this limit. Two approaches were tried which gave similar results. The integrand in (\ref{eq:I}) for $\omega\gg\mu$ becomes
\begin{equation}
{2\over\pi} (1+r)^2 \left[ {\Delta D^{(+)}\over k^2} - {(\lambda_2-\lambda_0) \over k^2} \right] \,,
\end{equation}
where
\begin{equation}
\Delta D^{(+)} = \Im D^{(+)} (\omega, 2\mu^2) - \Im D^{(+)} (\omega, 0) \,.
\end{equation}
The $(\lambda_2-\lambda_0)$ part is very small because of (\ref{eq:lambda0}) and (\ref{eq:lambda2}) and will be ignored.

We make two estimates for $\Delta D^{(+)}$ in the asymptotic region. In the first we use the phase shift analysis\cite{arndt} to give $\Delta D^{(+)}(\omega_0)$ at $\omega_0 \sim 2$~GeV. Then the asymptotic contribution is assume to be $\Delta D^{(+)}(\omega) = {\omega\over\omega_0} \Delta D^{(+)}(\omega_0)$. This would be expected for diffractive (Pomeron) dominance. Alternatively we used a comprehensive Regge pole analysis\cite{bp} to provide $\Delta D^{(+)}$. This approach incorporates $P$, $P'$ and $P''$ Regge exchanges. Both estimates give
\begin{equation}
I_{\rm ASY} \simeq 0.01 \,,
\end{equation}
which has been incorporated into the result (\ref{eq:continuum}).

We next consider the nucleon Born term in (\ref{eq:final}) which, with $g^2/4\pi\simeq 14$, largely cancels the continuum part (\ref{eq:continuum}) above to give
\begin{equation}
{\mu g^2\over M} \left(r\over 1-r^2\right)^2 - I \simeq -0.06\pm 0.02\,. \label{eq:result1}
\end{equation}

As a final step we separate out the small contributions of the $\ell\geq2$ scattering lengths. We use the  Koch dispersive analysis\cite{koch} for these higher scattering lengths to write (\ref{eq:goo}) as\cite{check}
\begin{equation}
D^{(+)}(\nu_T, 2\mu^2) = 4\pi \left( (1+r)^2 a_{0+}^{(+)} + a_{1-}^{(+)} + \left[ 3(1+r)^2 -1\right]a_{1+}^{(+)} \right) - (0.02\pm0.02) \,. \label{eq:result2}
\end{equation}
Here and subsequently we assume threshold parameters are in charged pion mass units.
Combining the results (\ref{eq:boo}), (\ref{eq:result1}) and (\ref{eq:result2}) with our general sum rule (\ref{eq:final}) yields
\begin{eqnarray}
{1\over4\pi} \overline D(2\mu^2) &=& (1+r)^2 \left[ a_{0+}^{(+)} - {1.047\over 3} \left( a_{0+}^{(1/2)} \right)^2 - {2\over3} (1.047) \left( a_{0+}^{(3/2)} \right)^2 - {1\over3} (1+2r)C^{(+)} \right] \nonumber\\
&& {}- {r^2\over 1 + 2r} a_{1-}^{(+)} + {r\over 1 + 2r} (6 + 13r + 6r^2) a_{1+}^{(+)} - {(0.08\pm 0.03)\over 4\pi} \,. 
\end{eqnarray}
Using the value $r = 0.0744$ from (\ref{eq:r}) we find (in units of $\mu$)
\begin{eqnarray}
\overline D(2\mu^2) &=& 14.5 a_{0+}^{(+)} - 5.06 \left( a_{0+}^{(1/2)} \right)^2 - 10.13 \left( a_{0+}^{(3/2)} \right)^2 - 5.55 C^{(+)} \nonumber\\
&& \hskip.6in {}- 0.06 a_{1-}^{(+)} + 5.70 a_{1+}^{(+)} - (0.08\pm0.03) \,. \label{eq:last}
\end{eqnarray}
We see that if the $s$-wave and $p$-wave scattering lengths and the effective range $C^{(+)}$ are known the sigma term $\sigma(2\mu)$ then follows by (\ref{eq:sigmaNN}).
We note that the last term in (\ref{eq:last}), which represents the net contributions of Born, continuum, and higher partial wave scattering lengths, is small. The near cancellation of the coefficient of $a_{1-}^{(+)}$ also makes it irrelevant. It turns out that $a_{1+}^{(+)}$ and $C^{(+)}$ are the dominant terms. Although $a_{0+}^{(+)}$ does not make a large contribution due to its small magnitude, it must be accurately determined because of its large weighting.

A recent analysis \cite{woolcock} parametrizes the very low energy $\pi^\pm p$ data with a $K$-matrix expansion. Applying this method  yields the $s$- and $p$-wave scattering lengths as well as the $s$-wave effective range parameters $C^{(\pm)}$. The threshold parameters which are needed in our sum rule (\ref{eq:last}) are given\cite{woolcock} (private communication):
%
\begin{eqnarray}
a_{0+}^{(+)} &=& \phantom+0.0035\pm0.0026\,,\\
a_{0+}^{(1/2)} &=& \phantom+0.179\pm0.006\,,\\
a_{0+}^{(3/2)} &=& -0.079\pm 0.003\,,\\
C^{(+)} &=& -0.116\pm0.026\,,\\
a_{1-}^{(+)} &=& -0.063\pm0.003,\\
a_{1+}^{(+)} &=& \phantom+0.135\pm0.002\,.
\end{eqnarray}
Inserting these into (\ref{eq:last}) and using (\ref{eq:sigmaNN}) we obtain
\begin{equation}
\sigma(2\mu^2) = 71 \pm 9\rm\ MeV \,.  \label{eq:result3}
\end{equation}
The dominant error is from the effective range $C^{(+)}$.

To show the results of earlier sets of threshold parameters, we first consider the very old set of Hamilton and Woolcock\cite{ham}. This set gives $\sigma_{\rm HW} (2\mu^2) = 64\pm9$~MeV. A slightly more recent set due to Koch\cite{koch} yields $\sigma_{\rm K} (2\mu^2) =55\pm6$~MeV\cite{last}. The slightly smaller values of $\sigma_{\rm HW}$ and $\sigma_{\rm K}$ can be ascribed to smaller values of $C^{(+)}$ and the change in sign of $a_{0+}^{(+)}$.

\section{Conclusions}

Combining subtracted $t=2\mu^2$ and $t=0$ dispersion relations allows the amplitude $\overline D(2\mu^2)$ to be determined by threshold parameters alone. By the ``on-shell theorem" (\ref{eq:sigmaNN}) the sigma term $\sigma(2\mu)$ then follows. The method presented here is an improvement on that used in \cite{olsson}. The central results obtained here are:

\begin{enumerate}

\item We demonstrate the relation of the sigma term $\sigma(2\mu)$ to threshold parameters.

\item Recent data and analyses\cite{woolcock} determine the scattering lengths and crossing-even effective range using only data near to threshold $(T_\pi < 100$~MeV). We use these results to compute $\sigma(2\mu^2) = 71\pm9$~MeV.
The assignment of error is straightforward using (\ref{eq:last}). This is in contrast to previous determinations using a global phase shift analysis with dispersive constraints where the error is difficult to determine. 

\item The value of $\sigma(2\mu^2)$ is not sensitive to $g_{\pi NN}$ but only depends on quantities close to the observables.

\item Our threshold method finds $\sigma(2\mu^2)$ in agreement with recent global calcuations\cite{gmor}.
The discrepancy between experiment and conventional ChPT theory\cite{gasser} is reinforced. 

\end{enumerate}

\section*{Acknowledgments}
This research was supported in part by the U.S.~Department of Energy under Grant No.~DE-FG02-95ER40896 and in part by the University of Wisconsin Research Committee with funds granted by the Wisconsin Alumni Research Foundation.


\newpage

\begin{figure}
\centering\leavevmode
\epsfxsize=5in\epsffile{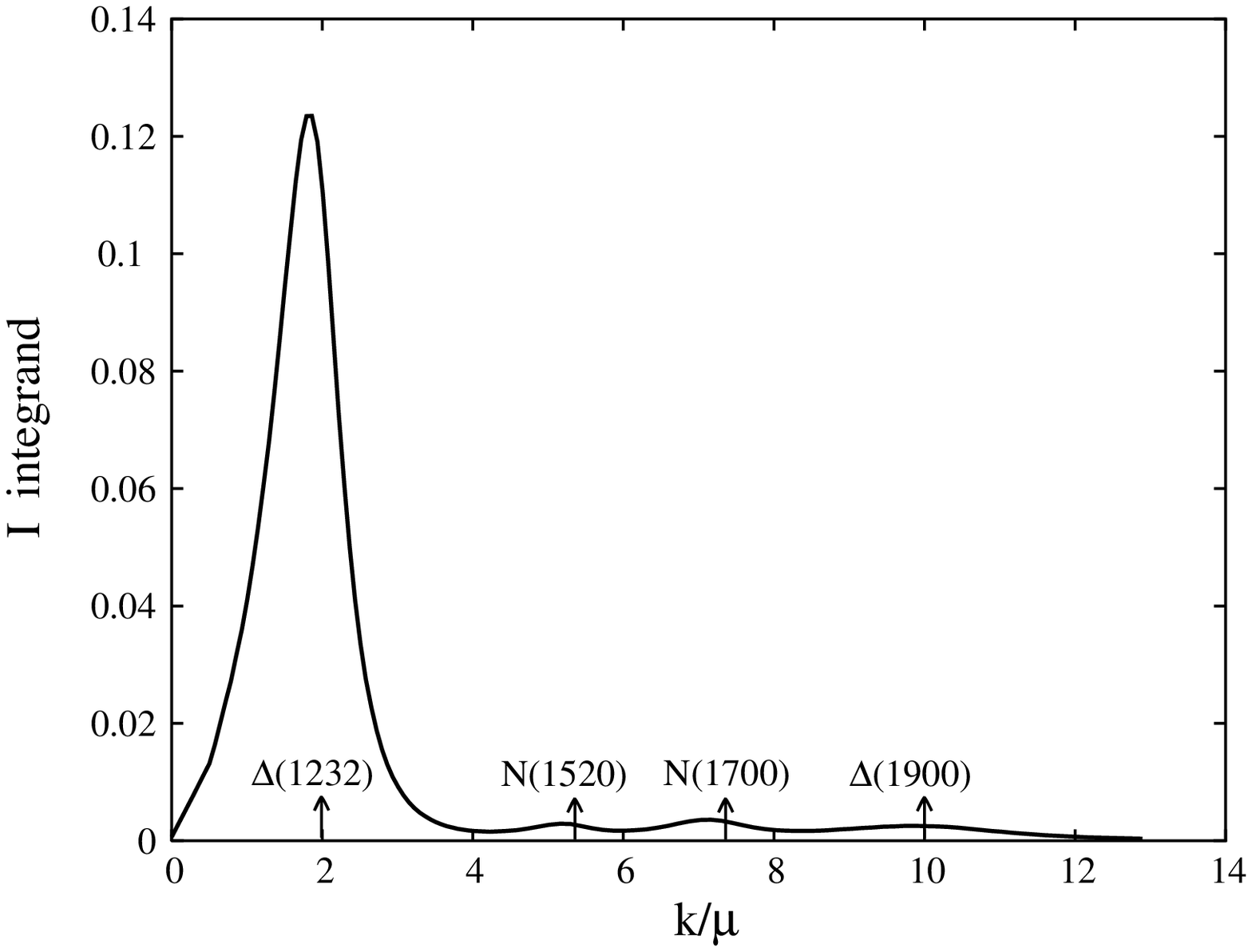}

\bigskip
\caption[]{The integrand (\ref{eq:I}) of the sum rule (\ref{eq:final}) showing the continuum contributions. Large cancellations occur between the $t=2\mu^2$ and $t=0$ parts of the integrand yielding a small result dominated by the $\Delta(1232)$ state. The evaluation employed the lowest thirteen partial waves of the VPI/GW phase shift analysis\cite{arndt}.}
\end{figure}


\begin{references}
\bibitem{eq1.4} In Eq.~(\ref{eq:amp}) the square bracket could be replaced by $\left[ 1 + {q^2 f(q^2) + q'^2 f(q'^2) \over \mu^2} \right]$ with no change as long as $q^2 f(q^2)\to 0$ for $q\to0$ and $f(-\mu^2)\equiv 1$. This could incorporate ``chiral logs".
\bibitem{coeff} The numerical coefficient 1.047 results from the numerical integration of (\ref{eq:F2mu}) with the replacement (\ref{eq:lim}). In the limit $r\to0$ this coefficient is unity.
\bibitem{check} To check further the convergence of (\ref{eq:result2}) we note that as $\ell$ increases, $\beta_{\ell+1, \pm} / \beta_{\ell,\pm}\to 2$, whereas $a_{\ell+1,\pm}/a_{\ell,\pm}\simeq 1/5$ according to Koch\cite{koch} for $\ell$ out to 10.
\bibitem{last} Koch and H\"ohler define the effective range as $\Re  f_{0+}^{(+)} = a_{0+}^{(+)} + b_{0+}^{(+)} Q^2$. Comparing with Hamilton and Woolcock's $C^{(+)}$ as defined in (\ref{eq:expansion}) we see that ${1\over3}(1+2r)^2 C^{(+)} = b^{(+)}_{0+} +{a^{(+)}_{0+}\over 2M\mu}$. The resulting $C^{(+)} = -0.103\pm0.015$.
\end{references}
\end{document}